\documentclass[aps,groupedaddress,superscriptaddress,prd,nofootinbib,twocolumn]{revtex4}

\usepackage{amsmath,amssymb,bm}
\usepackage{graphicx}
\usepackage{epstopdf}
\usepackage{amsfonts}
\usepackage{amssymb}
\usepackage{amsbsy}
\usepackage{amsmath}
\usepackage{latexsym}
\usepackage{sansmath}
\usepackage{subfigure}
\usepackage{lipsum}
\usepackage{float}
\usepackage{bm}
\usepackage{color}
\usepackage{comment}
\usepackage{csquotes}								
\usepackage{physics}
\usepackage{accents}

\usepackage[plain]{algorithm}
\usepackage{algpseudocode}

\def\be{\begin{equation}}
\def\ee{\end{equation}}
\def\bea {\begin{eqnarray}}
\def\eea {\end{eqnarray}}
\def\nn {\nonumber}

\begin{document}
 
\title{Dynamics and entanglement in quantum and quantum-classical systems:\\ lessons for gravity}

\author{Viqar Husain} \email{vhusain@unb.ca} 
\affiliation{Department of Mathematics and Statistics, University of New Brunswick, Fredericton, NB, Canada E3B 5A3}
\author{Irfan Javed} \email{i.javed@unb.ca}
\affiliation{Department of Mathematics and Statistics, University of New Brunswick, Fredericton, NB, Canada E3B 5A3}
 \author{Suprit Singh} \email{suprit@iitd.ac.in} 
\affiliation{Department of Physics, Indian Institute of Technology Delhi, Hauz Khas, New Delhi, India 110016} 

\begin{abstract}
\vskip 0.2cm
Motivated by quantum gravity, semi-classical theory, and quantum theory on curved spacetimes, we study the system of an oscillator coupled to two spin-1/2 particles. This model provides a prototype for comparing three types of dynamics: the full quantum theory, the classical oscillator with spin backreaction, and spins propagating on a fixed oscillator background. From nonperturbative calculations of oscillator and entanglement entropy dynamics, we find that entangled tripartite states produce novel oscillator trajectories, and that the three systems give equivalent dynamics for sufficiently weak oscillator-spin couplings, but deviate significantly for intermediate couplings. These results  suggest that semiclassical dynamics with back reaction does not provide a suitable  intermediate regime between quantum gravity and quantum theory on curved spacetime.   

\end{abstract}

\maketitle

A quantum theory of gravity (QG) is expected to provide a unification of gravity with the other forces of nature (for a recent review, see e.g. \cite{Carlip:2017dtj}).  The literature is abound with attempts to quantize gravity or simplified models of it \cite{Misner:1969hg} with no clear consensus so far on the approach to a final theory. If QG turns out to be a conventional quantum theory, it will be a system with a Hilbert space  ${\cal H} = {\cal H}_{\rm{gravity}}\otimes {\cal H}_{\rm{matter}}$. The matter component is in general  a ``multipartite" system representing several species of matter. This means that quantum states can have matter-gravity entanglement, and the corresponding entanglement entropy would be an evolving observable. 

If a QG theory were available, there would be several questions to pose.  The Universe we observe is well described by quantum fields on either a background of an expanding cosmology on large scales or a flat spacetime on smaller scales. One of the important questions is how such an approximation emerges dynamically from quantum gravity \cite{Kiefer:1993fg,Padmanabhan:2019art}. In between quantum gravity and quantum fields on a classical background spacetime, there is the intermediate regime of classical gravity coupled to quantum matter with backreaction. A proposal for this intermediate regime is the much studied semiclassical Einstein equation \cite{Singh:1989aa,Brout:1995aa,Anderson:1995tn}:
\be
G_{ab}(g) = 8\pi G\bra{\Psi}\hat{T}_{ab}\left(g,\hat{\phi}\right)\ket{\Psi}.
\ee
If this equation can be properly defined and solved, it would provide an association $(g,|\Psi\rangle)$ of a quantum state of matter with a classical metric $g$ (viewed in the Heisenberg picture). This is a nonperturbative hybrid classical-quantum equation; it raises many questions, such as what is the physical interpretation of the metric corresponding to a linear or entangled combination of matter states, and how exactly the right-hand side is to be defined if the metric is not known explicitly \cite{Isham:1995wr}. There are other hybrid models of this type: the so-called Newton-Schrodinger equation \cite{Diosi:2014ura,Bahrami:2014gwa}, a Friedmann-Schrodinger generalization to cosmology and related work \cite{Husain:2018fzg,Bojowald:2020emy}, and linear state evolution models using generalizations of the Lindblad equation \cite{Oppenheim:2018igd}.  

In a gravity-matter system it is of interest to study and compare three types of dynamics. These are the full quantum evolution, a suitably defined hybrid quantum-classical evolution with back reaction, and quantum evolution with no back reaction on the classical system. In the weak gravity regime, the gravitational field may be viewed as ``heavy" and slowly varying, and weakly coupled to much lighter and faster moving matter. In this regime of couplings, it is natural to expect that gravity behaves classically. On the other hand, in the deep QG regime, matter-gravity coupling would be strong and could produce highly entangled states. While a study of such comparative dynamics is technically challenging at the field theoretic level, it is relatively accessible in simpler models of gravity, such as cosmologies coupled to matter and non-gravitational systems.   

In this paper we study this set of questions in a model that has been a mainstay for work in atomic physics and quantum optics, the system of an oscillator coupled to a particle with spin, known as the Jaynes-Cummings model. We consider a slightly more general  model of an oscillator coupled to two spins-1/2 particles, together with a spin-spin coupling. In addition to studying the fully quantum case, we utilize the model in a new way by defining a coupled classical-quantum model with spin back reaction on the oscillator, and another without back reaction where the two spins propagate on an ``oscillator background."  The former case resembles a Hamiltonian version of the semi-classical Einstein equation whereas the latter may be viewed as simple case of quantum theory on curved spacetime. We study the comparative dynamics  numerically for a variety of initial states in the quantum-quantum (QQ) case and compare it with the dynamics in the semiclassical (SC) and classical background (CB) cases. (A related hybrid model with different dynamics is studied in \cite{fratino2014entanglement}.) 
Our main results are that the oscillator behaves classically even for highly entangled states in the QQ case for sufficiently weak oscillator-spin couplings, that initial product spin states can become maximally entangled in the SC and CB cases, and that the SC case has unusual static solutions not present in the other cases. We discuss implications of these results for gravitational systems, and for recent experimental proposals that dynamically generated entanglement of matter states may provide proof that linearized gravity must be quantum \cite{Bose:2017nin,Marletto:2017kzi,Marshman:2019sne}.

\begin{figure}
     \centering
     \includegraphics[width=0.6\columnwidth]{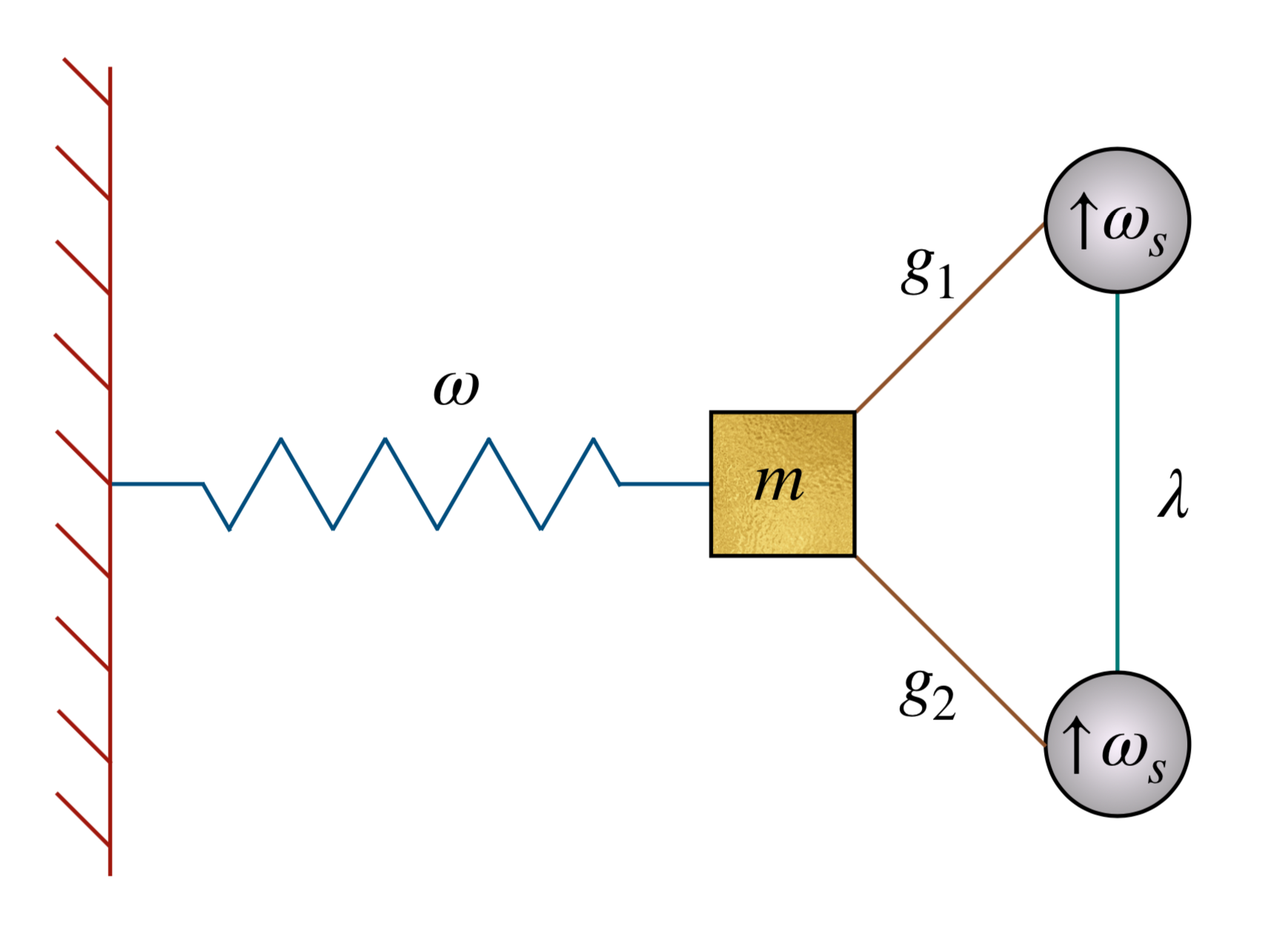}
     \caption{Oscillator coupled to two spin-$\frac{1}{2}$ particles.}
     \label{Fig0}
 \end{figure}
The system we consider is shown schematically in Fig.~\ref{Fig0}. The oscillator takes the place of gravity, and the spins correspond to matter. The first case is quantum-quantum, where the entire system is quantized; the second is coupled classical oscillator-spin system with back reaction; and the third is (quantum) spin dynamics on a fixed classical oscillator background. (Generalizations of such models to gravity may be achieved by extending e.g. the scalar-cosmology case discussed recently \cite{Husain:2020uac}.) 

\noindent \underbar{Quantum oscillator-spin (QQ)} The Hilbert space of the model for this case is the tensor product of the individual Hilbert spaces of the oscillator and the two spins, $\displaystyle
{\cal H} = {\cal H}_{o}\otimes {\cal H}_{\frac{1}{2}}^{(1)}\otimes {\cal H}_{\frac{1}{2}}^{(2)}$,
and the Hamiltonian is
\begin{widetext}
\begin{equation}
\begin{aligned}[b]
H = &\:\left(\frac{p^2}{2m}+\frac{1}{2}m\omega^2x^2\right)\otimes\left(I^{(1)}\otimes I^{(2)}\right)+I\otimes\frac{\omega_S}{2}\left(\sigma^{(1)}_z\otimes I^{(2)}+I^{(1)}\otimes \sigma_z^{(2)}\right)
+\frac{g_1}{2}\left(a\otimes\sigma^{(1)}_++a^{\dagger}\otimes\sigma_-^{(1)}\right)\otimes I^{(2)}\\
&\:+\frac{g_2}{2}\left(a\otimes I^{(1)}\otimes\sigma^{(2)}_++a^{\dagger}\otimes I^{(1)}\otimes\sigma_-^{(2)}\right)
+\frac{\lambda}{2}I\otimes\left(\sigma^{(1)}_+\otimes\sigma_-^{(2)}+\sigma^{(1)}_-\otimes\sigma_+^{(2)}\right)
\equiv h_o + h_s + h_{os} + h_{ss},
\end{aligned}
\label{HQQ}
\end{equation}
\end{widetext}
where $\sigma_z,\sigma_\pm$ are the Pauli diagonal and ladder operators, and $a= x\sqrt{m\omega/2} +  ip/\sqrt{2m\omega}$. The first two terms in Eq.~(\ref{HQQ}) are  the Hamiltonians of the noninteracting oscillator and spins ($h_{o}$ and $h_{s}$), the second two are the interactions of the oscillator with each of the spins with coupling constants $g_1/2$ and $g_2/2$ ($h_{os}$), and the third is the spin-spin interaction with coupling $\lambda/2$ ($h_{ss}$).

We restrict our attention to the $4\times 4$ truncation of the oscillator Hamiltonian and consider initial states that are linear combinations of the ground and first excited states. This ensures that the coupled quantum dynamics remains in the $16$-dimensional Hilbert space $\mathcal{H} = \mathcal{H}_{o}\otimes\mathcal{H}^{(1)}_{\frac{1}{2}}\otimes\mathcal{H}^{(2)}_{\frac{1}{2}}$.  Thus, the time dependent Schrodinger equation (TDSE) in this truncation is a set of $16$ coupled ODEs.

\noindent\underbar{Semiclassical oscillator-spin (SC)} In this case the oscillator is classical with orbits in the $\mathbb{R}^2$ phase space with coordinates $(x, p)$, and the spin state is given by a vector in the Hilbert space ${\cal H}_{\frac{1}{2}}^{(1)}\otimes {\cal H}_{\frac{1}{2}}^{(2)}$. The Hamiltonian is
\begin{widetext}
\begin{equation}
\begin{aligned}[b]
H = &\:\left(\frac{p^2}{2m}+\frac{1}{2}m\omega^2x^2\right)\left(I^{(1)}\otimes I^{(2)}\right)+\frac{\omega_S}{2}\left(\sigma^{(1)}_z\otimes I^{(2)}+I^{(1)}\otimes \sigma_z^{(2)}\right) +\frac{g_1}{2}\left(a\sigma^{(1)}_++a^*\sigma_-^{(1)}\right)\otimes I^{(2)}\\ &\:+I^{(1)}\otimes \frac{g_2}{2}\left(a\sigma^{(2)}_++a^*\sigma_-^{(2)}\right)+\frac{\lambda}{2}\left(\sigma^{(1)}_+\otimes\sigma_-^{(2)}+\sigma^{(1)}_-\otimes\sigma_+^{(2)}\right)
\equiv h_o^{SC} + h_s^{SC} + h_{os}^{SC} + h_{ss}^{SC},
\end{aligned}
\label{HSC}
\end{equation}
\end{widetext}
 where each component is defined as for the fully quantum case. However $x$, $p$, $a$, and $a^*$ are now classical variables. We define the coupled dynamics with the TDSE for the spins and the Hamilton equations for the oscillator:  
\bea
i\frac{d}{dt}\ket{\Psi} &=& \left(h_s^{SC} +h_{os}^{SC} + h_{ss}^{SC}\right)\ket{\Psi}\\
   \dot{q} &=& \{q, H_{{\rm eff}}\}, \quad
   \dot{p} = \{p, H_{{\rm eff}}\}, 
  \eea
where $\ket{\Psi}$ is a spin state and 
\be
H_{\rm{eff}}(x,p) \equiv \langle \Psi| H|\Psi\rangle.
\ee
These are a set of $6$ coupled ODEs to be solved with initial data set $\{ x_0, p_0, |\psi\rangle_0\}$.  The quantum dynamics is unitary by definition, and it is readily verified using the evolution equations that the Hamiltonian $H_{{\rm eff}}$ is a constant of motion.  
  
\noindent\underbar{Spins on classical (oscillator) background (CB):} This case is the simplest of the three. We define it by fixing an classical oscillator ``background" solution $(x_c(t),p_c(t))$ and $a_c = x_c\sqrt{m\omega/2}+  ip_c/\sqrt{2m\omega}$ with parameters $m$ and $\omega$, and the time dependent spin Hamiltonian 
 \begin{widetext}
 \bea
H &=& \frac{\omega_{S}}{2} \left(\sigma^{(1)}_z\otimes I^{(2)} + I^{(1)}\otimes \sigma_z^{(2)}\right)  + \ \frac{g_1}{2}\ \left( a_{c}(t)\sigma^{(1)}_+ + a_{c}^*(t)\sigma_-^{(1)}  \right)\otimes  I^{(2)}   + 
 I^{(1)} \otimes \frac{g_2}{2}\left( a_{c}(t)\sigma^{(2)}_+ + a_{c}^*(t)\sigma_-^{(2)} \right)\nn\\
 &&+\ \frac{\lambda}{2}\  \left( \sigma^{(1)}_+\otimes \sigma_-^{(2)} + \sigma^{(1)}_- \otimes\sigma_+^{(2)}  \right).
  \eea
  \end{widetext}
  Dynamics is defined solely by the TDSE of the spin state. Thus this is a system of 4 coupled ODEs for the spin state.
 
 \noindent\underbar{Comparing dynamics} All three oscillator-spins cases defined above (QQ, SC, and SB) have dimensional parameters $m$, $\omega$, $\omega_S$, $g_1$, $g_2$, and $\lambda$. The oscillator provides ``fundamental"  time and length scales  $1/\omega$ and $1/\sqrt{m\omega}$, respectively (with $\hbar =1$). We set these equal to unity and measure the remaining four parameters in these units.
 
 Comparing dynamics in the three systems is accomplished by first  fixing initial data  $\{x(0)$, $p(0),|\Psi\rangle_s(0)\}$ for the SC system, and then (i) using the same initial spin state $|\Psi\rangle_s(0)$ for the QQ and CB systems, and (ii) matching initial data for the oscillator. The latter is accomplished for the CB case by using the oscillator solution that goes through the phase space point $(x(0), p(0))$, and for the QQ case by using the product oscillator-spin state
 \begin{equation}
  | \Phi\rangle(0)= \left(\cos\left(\frac{\theta}{2}\right)|0\rangle + \sin\left(\frac{\theta}{2}\right)e^{i\phi} |1\rangle\right)\otimes  |\Psi\rangle_s(0), \label{QQState} 
  \end{equation}
 (where $|0\rangle$ and $|1\rangle$ are respectively the ground and first excited states of the oscillator), and then fixing $\theta$ and $\phi$ such that the expectation values of $\hat{x}$ and $\hat{p}$ match: 
 \bea
 x(0) &=& 
 \frac{1}{\sqrt{2m\omega}}\sin(\theta)\cos(\phi)
 \label{state_match_1}\\
p(0) &=&    -\sqrt{\frac{m\omega}{2}}\sin(\theta)\sin(\phi).
 \label{state_match_2}
 \eea
 This ensures the closest possible initial data for all the three cases so that we can compare $(x, p)$ and $(\langle \hat{x}\rangle, \langle \hat{p}\rangle)$, phase space trajectories, and the evolution of spin entanglement entropy and energy in the oscillator and spin subsystems. There is also the interesting possibility of considering the maximally entangled Greenberger–Horne–Zeilinger (GHZ) states of the QQ system as initial states and computing the phase space plots $(\langle\hat{x}\rangle, \langle\hat{p}\rangle)$. However, there is no obvious comparison that can be  made with dynamics in the SC and SB systems.

 
 \begin{figure*}
     \includegraphics[width=\textwidth]{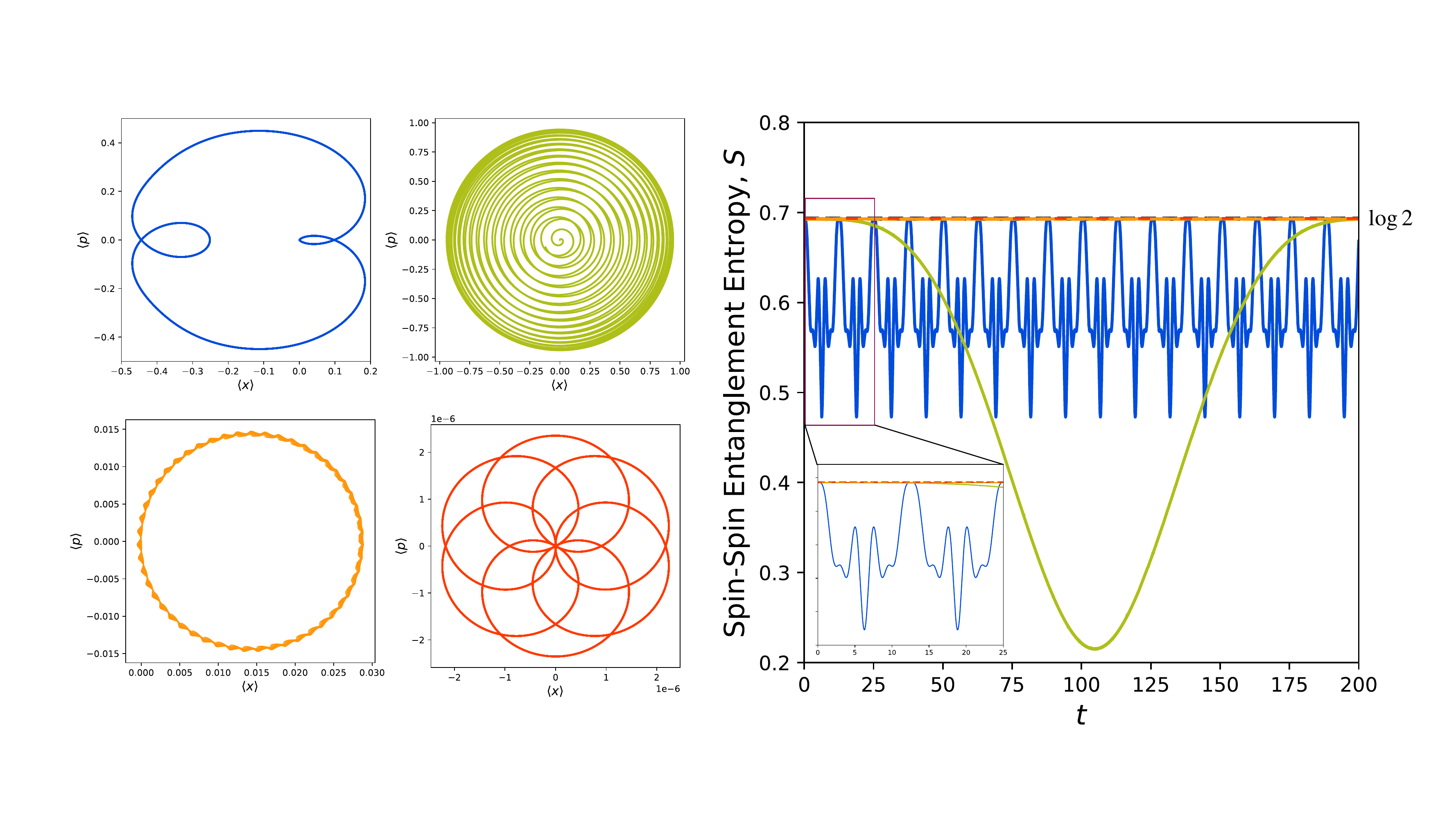}
     \caption{QQ system for initial GHZ state with $m=\omega=1$. The $2\times 2$ square shows phase space plots $(\langle \hat{p} \rangle, \langle \hat{x} \rangle)$: top left is for $\omega_S=g_1=g_2=\lambda=1$; top right for $\omega_S=g_1=g_2=1$ and $\lambda=100$; bottom left for $\omega_S=0.5$, $g_1=g_2=1$, and $\lambda=100$; and bottom right for $\omega_S=4$, $g_1=g_2=0.1$, and $\lambda=2000$. These illustrate the variety of oscillator phase space dynamics;  the last case is typical of the intricate  phase space patterns that arise for large spin-spin coupling. The frame on the right shows the corresponding spin entanglement entropy as a function of time; it is constant at $\log 2$ for cases in the second row, but shows an oscillatory pattern for the top left case and a slow dip and rise for the top right case. (The black dot is the initial value of $(\langle \hat{p} \rangle, \langle \hat{x} \rangle)$ fixed by the initial GHZ state; it is the same for all cases shown.)}
     \label{ghz-plot}
 \end{figure*}
 
 For all cases we integrated the coupled differential equations numerically for a variety of initial data. The method we used ensured that probability and $H_{\rm eff}$ are conserved  at least to order $10^{-8}$. Our first set of results is for the QQ system with the GHZ initial state 
 \begin{equation}
     \ket{\Phi}(0)= \frac{1}{\sqrt{2}}\left(|0 ++\rangle + |1 --\rangle \right).
     \label{ghz}
 \end{equation}
 Fig.~\ref{ghz-plot} shows phase space plots $(\langle \hat{p}\rangle,\ \langle \hat{x} \rangle)$ and entanglement entropy evolution for a selection of parameters. The former provide a remarkable characterization of the oscillator behaviour. These cannot be compared with the other two cases, where the oscillator remains classical, but may provide unique signatures of the GHZ state: for instance, the bottom left frame of Fig.~\ref{ghz-plot} is a fuzzy ellipse that is not centered at the origin, a feature that does not arise in the SC and CB cases.

\begin{figure*}
     \centering
     \includegraphics[width=0.9\textwidth]{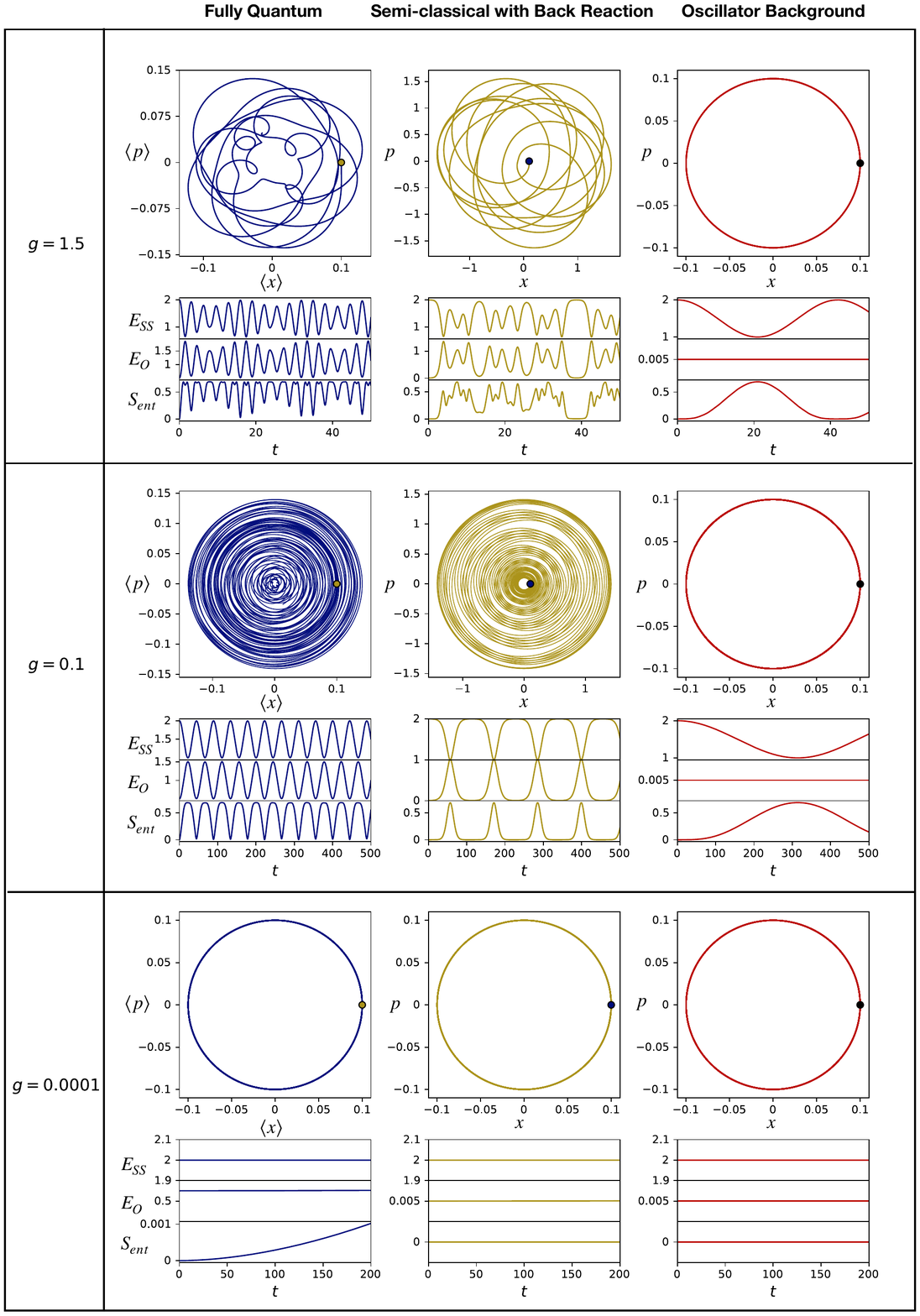}
     \caption{Phase space, spin entanglement entropy $S_{\rm ent}$, and subsystem energies $E_{\rm osc}$ and $E_{\rm ss}$ for $m = \omega = 1$ and $\omega_{S} =\lambda = 2$ for the $g$ values indicated. The initial data for the  semiclassical cases is $x = 0.1$, $p = 0$ (indicated with a black dot), and spin state $\ket{++}$;  the corresponding data for the fully quantum case from (\ref{QQState}) is  $(0.99748420879\ket{0}+0.07088902028\ket{1})\ket{++}$. (The phase space plot axes represent $\langle x\rangle$ and $\langle p\rangle$ for the QQ, and $x,p$ for the SC and CB cases.)}
     \label{3cases}
 \end{figure*}

 We computed several solutions of the three cases with comparable initial data (as described above) with the aim of studying the parameter ranges where the oscillator dynamics looks similar. A representative sample is shown in Fig.~\ref{3cases} with parameter values $m=\omega=1$, $\omega_S=\lambda=2$, and oscillator-spin coupling parameters $g\equiv g_1=g_2 = 0.0001, 0.1, 1.5$. The latter are chosen to highlight how the dynamics in phase space, spin entanglement entropy $S_{ent}$, and energies in the spin subsystem  $E_{\rm ss} = \langle h_s + h_{os} + h_{ss} \rangle$ and oscillator $E_{\rm osc} = \langle H\rangle - E_{\rm ss}$  change with oscillator-spin couplings.  
 
 The initial data for the SC and CB cases in  Fig. \ref{3cases} is $\{q(0)=0.1, p(0)=0, |\Psi(0)\rangle =|++\rangle\}$. The corresponding initial state for the QQ system is obtained by using Eqs.~(\ref{state_match_1}) and (\ref{state_match_2}) with $\phi=0$ and $\sin\theta = 0.1\sqrt{2}$; these are very close to the (truncated) oscillator coherent states. 
 
 We highlight the following features evident in Fig.~\ref{3cases}: (i) for $g=0.0001$, the oscillator phase space trajectories and subsystem energies are indistinguishable in all three cases, and the spin entanglement entropy remains nearly zero; (ii) as $g$ increases  to $0.1$, differences start to appear in each of the variables plotted, and in particular, spin entanglement entropy increases from zero, attains its maximum value of $\log 2$, and oscillates, notably {\it even} for the SC and CB cases; (iii) spin entanglement entropy and subsystem energy oscillations have higher frequencies for larger $g$ values; and (iv) the SC phase space trajectory is more expansive, with the range of the oscillator extending an order of magnitude more than that for the QQ and CB cases. Similar features are evident for other parameter values and initial data. 
 While it is gratifying to note that all cases approximately agree for sufficiently small $g$ values, point (iv) above is especially noteworthy: it provides evidence that the backreaction SC model, which is very similar in form and spirit to the semiclassical Einstein equation,  may not provide a reasonable transition between the QQ and CB systems.  
 
 The SC equations provide a further curious feature not present in the QQ and CB systems: static solutions for the oscillator. These are obtained by considering eigenstates of the spin subsystem $h_s+h_{os}+h_{ss}$, and setting $\dot{x}=\dot{p}=0$. The SC equations then reduce to
 \bea 
 \dot{x} &=&\frac{p}{m} + \{x,E(a,a^*,g,\lambda)\}=0\\
 \dot{p} &=&-m\omega^2 x + \{p,E(a,a^*,g,\lambda)\}=0,
 \eea
where $E(a,a^*,g,\lambda)$ are the corresponding eigenvalues. For $\lambda=0$ there are particularly simple static solutions: any point on the circle $x^2 + p^2 = g^2/2 - 2\omega_s^2/g^2$ with $m\omega=1$. Thus we must have $g^2>2\omega_s$. The physical interpretation of  the solution with $p=0, x\ne0$ is that the stationary spin state ``holds" the stretched spring of the oscillator. But the physical interpretation of solutions with $x\ne 0,\ p\ne 0$  are unusual: the spring is held stretched from equilibrium (since $x\ne 0$) and the mass is in uniform motion since $p\ne 0$! Other static solutions are readily computed numerically.

\noindent\underbar{Discussion} We described in detail three versions of the dynamics of the oscillator coupled to two spin-$1/2$ particles (QQ, SC, and CB) with a truncation of the oscillator to a 4-level system.  Our aim was to compare the dynamics of the oscillator, spin entanglement entropy, and subsystem energies for the same initial conditions. We highlight three results and comment on some implications:
  \begin{itemize}
      \item For sufficiently small oscillator-spin couplings, the dynamics of the three systems is identical; this lends support to the idea that similar results would hold for other systems, including gravity coupled to matter.
      \item Of particular note is that the SC system gives oscillator trajectories that are substantially different from the QQ system for larger oscillator-spin couplings. This may be attributed to the fact that the SC equations are non-linear in the state, unlike the QQ and CB systems. As the same  (but more consequential) nonlinearity holds for the semiclassical Einstein equation, our results suggest that the latter does not provide the appropriate transition between quantum gravity and quantum fields on curved spacetime. 
      
      \item Spin entanglement is induced in the SC and CB systems for non-zero spin-spin coupling $\lambda$; e.g. $g=0.1$, $\lambda =2$ in Fig.\ref{3cases}. (The CB case is similar to entanglement generation in Floquet dynamics \cite{claeys2018spin}.) The implication for gravity is similar: initial product states of matter can get entangled thorough the semiclassical Einstein equation, or even by propagating on a fixed but time dependent background spacetime, provided matter is self-interacting through any local field. 
      \item In the proposed experiments \cite{Bose:2017nin,Marletto:2017kzi,Marshman:2019sne} for detecting quantization of linearized gravity through entanglement generation in mass states, it is posited that interaction between masses is not action-a-distance (as it is here for spin-spin), but is instead generated via a mediating quantum gravitational field. However, if the masses are sufficiently close in a laboratory setting, a point interaction may be a good approximation, and any entanglement generated through non-gravitational quantum interactions, whether local or not, could be significant as demonstrated in the model discussed here. (See Ref. \cite{Fragkos:2022tbm} for related discussion.) In the final analysis, a quantum interaction is of course necessary to generate entanglement between masses, what ever its origin, and the spin-spin interaction in the present model is a stand-in for that.
      
      \item There is a curious case for the SC model where for the $|++\rangle$ or $|--\rangle$ initial states, there is no entanglement induced for $g=0,\ \lambda\ne 0$. Then, increasing $g$ from zero (i.e. turning on the classical coupling) induces entanglement; this special case is an exception to the proof in \cite{Hall:2017nzl} which covers the $\lambda = 0$ case.
      
  \end{itemize}
 
 Our results suggest several areas for further investigation. These include considering in the same spirit cosmological and other gravitational models with scalar and/or spinorial fields by extending the work in \cite{Husain:2020uac}, a field theoretic version of the SC model  for studying back reaction induced entanglement in gravity coupled to a scalar field in spherically symmetric gravity where matter gravity entanglement is a potentially important feature \cite{Husain:2009vx}, and linear alternatives to the semiclassical Einstein equation as discussed in \cite{Oppenheim:2018igd} applied to similar model systems; the latter may address the issue of the significant difference between the QQ and SC systems for the results presented here.
 
 There are a few broader features of coupled classical-quantum systems where the quantum part evolves unitarily and the classical one via Hamilton equations. The SC equations are well defined as a dynamical system whether physically valid or not, and the corresponding effective Hamiltonian is conserved; the same holds for the semiclassical Einstein equation at least for model systems \cite{Husain:2018fzg}. This raises questions for the type of analysis presented in \cite{Maudlin:2019bje}, where the question of energy conservation is raised. Secondly, can classical-quantum systems, the way we have defined them here, be viewed as ``ongoing measurement" of a quantum system by a classical apparatus (here the oscillator)? Both the oscillator and the spins undergo continuous smooth evolution, and if the oscillator is  macroscopic, it would not itself be significantly disturbed by backreaction from the quantum system.
\smallskip

\noindent \underbar{Acknowledgements} This work was supported by the Natural Science and Engineering  Research Council of Canada. S.S. is supported in part by the Young Faculty Incentive Fellowship from IIT Delhi. We thank Stijn De Baerdemacker, Carlo Rovelli, Mustafa Saeed, Danny Terno, Edward Wilson-Ewing, and Nomaan X for constructive comments on the manuscript.

 \bibliography{Osc-spins}
   
  \end{document}